\documentclass[twocolumn,aps,showpacs]{revtex4}

\usepackage{amsmath}
\usepackage{amssymb}
\usepackage{graphicx}
\usepackage{dcolumn}

\begin{document}

\title{Microwave-induced magnetoresistance of two-dimensional electrons interacting with acoustic phonons}

\author{O. E. Raichev}
\affiliation{Institute of Semiconductor Physics, National Academy of Sciences of Ukraine, 
Prospekt Nauki 41, 03028, Kiev, Ukraine}
\date{\today}

\begin{abstract}
The influence of electron-phonon interaction on magnetotransport in two-dimensional 
electron systems under microwave irradiation is studied theoretically. Apart from 
the phonon-induced resistance oscillations which exist in the absence of microwaves,
the magnetoresistance of irradiated samples contains oscillating contributions due 
to electron scattering on both impurities and acoustic phonons. The contributions due 
to electron-phonon scattering are described as a result of the interference of 
phonon-induced and microwave-induced resistance oscillations. In addition, microwave 
heating of electrons leads to a special kind of phonon-induced oscillations. The 
relative strength of different contributions and their dependence on parameters are 
discussed. The interplay of numerous oscillating contributions suggests a peculiar 
magnetoresistance picture in high-mobility layers at the temperatures when 
electron-phonon scattering becomes important. 
\end{abstract}

\pacs{73.23.-b, 73.43.Qt, 73.50.Pz, 73.63.Hs}

\maketitle

\section{Introduction}

Studies of high-mobility two-dimensional (2D) electron systems in weak perpendicular 
magnetic fields $B$ reveal several kinds of magnetoresistance oscillations which are 
caused by the scattering-assisted electron transitions between different Landau levels 
and survive increase in temperature $T$, in contrast to Shubnikov-de Haas oscillations. Most 
of attention is paid to the microwave-induced resistance oscillations (MIRO) which appear under 
steady-state microwave (MW) irradiation of 2D electron gas.$^1$ The MIRO periodicity is 
determined by the ratio of the radiation frequency $\omega$ to the cyclotron frequency 
$\omega_{c}$, while the amplitude is governed by the radiation power. As the power 
increases, MIRO minima are spread into intervals of zero dissipative resistance known 
as "zero-resistance states".$^{2,3,4}$ The basic theoretical consideration$^{5-9}$ of 
MIRO involves the picture of transitions between broadened Landau levels under elastic 
scattering of electrons by impurities or other inhomogeneities in the presence of 
MW excitation, when electrons absorb or emit radiation quanta. As a consequence 
of these transitions, both the electron scattering rate and the electron distribution 
function acquire oscillating components which lead to magnetoresistance oscillations. 
The corresponding contributions are often described in terms of "displacement"$^{5,6,7}$ 
and "inelastic"$^{8,9}$ microscopic mechanisms of MIRO, respectively. The above terminology 
emphasizes that the scattering-assisted MW absorption gives rise to spatial 
displacement of electrons along the applied dc field, while the modification of the 
distribution function is controlled by inelastic relaxation of electrons. Consideration 
of these mechanisms satisfactory explains both the periodicity and the phase of MIRO 
observed in experiments, as well as their power and temperature dependence.$^{10,11,12}$

Inelastic scattering of electrons by acoustic phonons makes possible the phonon-assisted 
transitions of electrons between Landau levels. The probability of such transitions is 
maximal when the phonon momentum is close to the Fermi circle diameter, $2p_F$. This 
leads to a special kind of magnetophonon oscillations also known as phonon-induced 
resistance oscillations (PIRO),$^{13-17}$ whose periodicity is determined by 
commensurability of the characteristic phonon energy with cyclotron energy. In 
GaAs quantum wells with very high mobility ($ \sim 10^7$ cm$^2$/V s), these 
oscillations are visible already at $T=2$ K,$^{17}$ while in the samples with moderate 
mobilities ($ \sim 10^6$ cm$^2$/V s) they are well seen at $T=10-20$ K.$^{15}$ A theoretical 
description of PIRO assuming interaction of 2D electrons with anisotropic bulk phonon 
modes and showing a good agreement with experimental data has been presented recently.$^{18}$

Both MIRO and PIRO are observed in the same interval of magnetic fields, $B < 1$ T, 
and have comparable periods. Therefore, it is interesting to study the interplay of these 
remarkable oscillating phenomena. In other words, one may pose a question: what happens 
with magnetoresistance of the sample which shows PIRO if this sample is irradiated 
by microwaves? More generally, there exists an important problem of the influence of 
electron-phonon interaction on MW-induced magnetoresistance. Some aspects of this 
problem have been considered previously. First, it is well established that increasing 
MW power leads to heating of the electron system, especially in the samples with
lower mobility. The effective electron temperature $T_e$ is determined by the power  
transmitted from electron system to the lattice via electron-phonon interaction.$^{19}$
The heating causes a suppression of MIRO amplitudes$^{11,20,21}$ for two main reasons: 
the enhancement of Landau level broadening and a decrease of the inelastic scattering time. 
In this sense, a consideration of electron-phonon interaction becomes necessary for 
description of MW photoresistance at elevated MW power. Further, as shown 
theoretically,$^{22,23}$ there is a direct influence of electron-phonon scattering on 
the photoresistance: this scattering in the presence of MW excitation can give rise to 
pronounced magnetoresistance oscillations whose behavior is different from that of the 
case of impurity-assisted transport. A consideration of electron-phonon interaction as a 
possible reason for the absolute negative resistivity and zero-resistance states 
has been also presented.$^{23}$ 

The aim of this paper is to give a consistent theoretical description of magnetotransport 
under MW irradiation in the presence of both elastic and phonon-assisted electron 
scattering. It is demonstrated that the magnetoresistance has contributions of MIRO, PIRO, 
and the terms corresponding to interference of these types of oscillations. Apart from this, 
heating of electron system leads to an additional PIRO contribution which is shifted by 
phase with respect to equilibrium PIRO. Superposition of all these contributions gives rise 
to a complicated oscillating resistivity picture which is essentially different from either 
equilibrium magnetoresistance or MW-modified magnetoresistance under elastic scattering only. 
The analytical expression for magnetoresistance is obtained by assuming weak excitation, when 
the response is linear in MW power. 

The paper is organized as follows. Section II describes the quantum Boltzmann 
equation for the system of 2D electrons interacting with impurities and phonons in the 
presence of microwaves. Section III is devoted to calculation of electron distribution 
function and resistivity. Section IV contains numerical results based on the expressions 
derived in Sec. III, description of different contributions to oscillating magnetoresistance, 
a discussion of the approximations used, and a brief conclusion.
 
\section{General formalism}

In the region of weak magnetic fields, when the cyclotron energy 
is small in comparison with the Fermi energy, it is convenient to calculate the 
resistivity by using the quantum Boltzmann equation for the generalized Wigner distribution 
function $f_{\varepsilon {\bf p} t}$, which depends on energy $\varepsilon$, quasiclassical 
2D momentum ${\bf p}$ and time $t$. The coordinate dependence of this function is 
omitted, since only spatially-homogeneous 2D systems are under consideration. To take 
into account the influence of MW radiation and dc excitation 
on 2D electrons, one may use a transition to the moving coordinate frame (see Ref. 9 
and references therein), when the MW and dc field potentials are transferred 
to the collision integrals and do not appear in the left-hand side of the kinetic 
equation. The scattering of electrons by impurities and phonons is treated within 
the self-consistent Born approximation (SCBA), which assumes that both the mean free path 
of electrons and the magnetic length are larger than the correlation lengths of the 
corresponding scattering potentials. Then, the quantum Boltzmann equation takes the form
\begin{equation}
\left(\frac{\partial}{\partial t} + \omega_c \frac{\partial}{\partial \varphi} \right) 
f_{\varepsilon {\bf p} t} = J^{im}_{\varepsilon {\bf p} t}+J^{ph}_{\varepsilon {\bf p} t}+ 
J^{ee}_{\varepsilon {\bf p} t},
\end{equation}
where $J^{im}$, $J^{ph}$, and $J^{ee}$ are the collision integrals for electron-impurity, 
electron-phonon, and electron-electron scattering, respectively. They are written as 
\begin{eqnarray}
J^{\kappa}_{\varepsilon {\bf p} t}=-i \int d \tau e^{i \varepsilon \tau} \biggl\{ \int d t' 
\left[\Sigma^{(\kappa)R}_{t+\tau/2,t'}({\bf p}) f_{t',t-\tau/2}({\bf p}) \right. \nonumber \\
\left. - f_{t+\tau/2,t'}({\bf p}) \Sigma^{(\kappa)A}_{t',t-\tau/2}({\bf p}) \right] - \Sigma^{(\kappa)-+}_{t+\tau/2,t-\tau/2}({\bf p}) \biggr\},
\end{eqnarray}
where $\kappa$ is the index of corresponding interaction ($im$, $ph$, or $ee$). Throughout the 
paper, the system of units with $\hbar=1$ is used. The distribution function in the double-time 
representation, $f_{t_1,t_2}({\bf p})$, is related to the Wigner distribution function 
$f_{\varepsilon {\bf p} t}$ by the Wigner transformation, which is defined for arbitrary 
function $A$ as $A_{\varepsilon {\bf p} t}= \int d \tau e^{i \varepsilon \tau} A_{t+\tau/2, 
t-\tau/2}({\bf p})$. The Keldysh self-energies 
are described by the following expressions:
\begin{equation}   
\Sigma^{(im)-+}_{t_1,t_2}({\bf p}) = - \int \frac{d {\bf q}}{(2 \pi)^2} w(q) e^{i {\bf q} \cdot {\bf R}_{t_1 t_2}}
G^{-+}_{t_1,t_2}({\bf p}-{\bf q}),
\end{equation}
\begin{eqnarray}   
\Sigma^{(ph)-+}_{t_1,t_2}({\bf p}) = -i \int \frac{d {\bf q}}{(2 \pi)^2} \int_{-\infty}^{\infty} \frac{d q_z}{2 \pi} 
\sum_{\lambda} M_{\lambda {\bf Q}} e^{i {\bf q} \cdot {\bf R}_{t_1 t_2}} \nonumber \\
\times \int \frac{d \Omega}{2 \pi} D^{-+}_{\lambda {\bf Q}}(\Omega) e^{-i \Omega (t_1-t_2)}
G^{-+}_{t_1,t_2}({\bf p}-{\bf q}),
\end{eqnarray}
and
\begin{eqnarray}   
\Sigma^{(ee)-+}_{t_1,t_2}({\bf p}) = - 2 \int \frac{d {\bf q}}{(2 \pi)^2} U^2_q G^{-+}_{t_1,t_2}({\bf p}+{\bf q})
\nonumber \\
\times \int \frac{d {\bf p}'}{(2 \pi)^2} G^{-+}_{t_1,t_2}({\bf p}'-{\bf q}) G^{+-}_{t_2,t_1}({\bf p}').
\end{eqnarray}
The retarded and advanced self-energies, $\Sigma^{(\kappa)R}$ and $\Sigma^{(\kappa)A}$, can be expressed as 
\begin{eqnarray}
\Sigma^{(\kappa)R} = \Sigma^{(\kappa)-+}+\Sigma^{(\kappa)--}, \nonumber \\
\Sigma^{(\kappa)A} = -\Sigma^{(\kappa)-+}- \Sigma^{(\kappa)++},
\end{eqnarray}
where
\begin{equation}   
\Sigma^{(im) \pm \pm}_{t_1,t_2}({\bf p}) = \int \frac{d {\bf q}}{(2 \pi)^2} w(q) e^{i {\bf q} \cdot {\bf R}_{t_1 t_2}}
G^{\pm \pm}_{t_1,t_2}({\bf p}-{\bf q}),
\end{equation}
\begin{eqnarray}   
\Sigma^{(ph) \pm \pm}_{t_1,t_2}({\bf p}) = i \int \frac{d {\bf q}}{(2 \pi)^2} \int_{-\infty}^{\infty} 
\frac{d q_z}{2 \pi} 
\sum_{\lambda} M_{\lambda {\bf Q}} e^{i {\bf q} \cdot {\bf R}_{t_1 t_2}} \nonumber \\
\times \int \frac{d \Omega}{2 \pi} D^{\pm \pm}_{\lambda {\bf Q}}(\Omega) e^{-i \Omega (t_1-t_2)}
G^{\pm \pm}_{t_1,t_2}({\bf p}-{\bf q}),
\end{eqnarray}
and
\begin{eqnarray}   
\Sigma^{(ee) \pm \pm}_{t_1,t_2}({\bf p}) = 2 \int \frac{d {\bf q}}{(2 \pi)^2} U^2_q G^{\pm \pm}_{t_1,
t_2}({\bf p}+{\bf q}) \nonumber \\
\times \int \frac{d {\bf p}'}{(2 \pi)^2} G^{\pm \pm}_{t_1,t_2}({\bf p}'-{\bf q}) G^{\pm \pm}_{t_2,t_1}({\bf p}').
\end{eqnarray}

The quantities entering Eqs. (3)-(9) are described below. First, 
$G^{-+}_{t_1,t_2}({\bf p})$ is the Keldysh Green's function for electrons, 
which is related to $f_{t_1,t_2}({\bf p})$ as
\begin{equation}   
G^{-+}_{t_1,t_2}({\bf p})= \int d t' \left[ f_{t_1,t'}({\bf p}) G^{A}_{t',t_2}({\bf p})- G^{R}_{t_1,t'}({\bf p}) 
f_{t',t_2}({\bf p}) \right],
\end{equation}
where $G^{A}$ and $G^{R}$ are the retarded and advanced Green's functions. The Green's 
function $G^{+-}$ is expressed by the linear combination, $G^{+-}=G^{-+}+G^{R}-G^{A}$. 
Next, $D^{-+}_{\lambda {\bf Q}}(\Omega)$ is the Keldysh Green's function for phonons: 
\begin{eqnarray}   
D^{-+}_{\lambda {\bf Q}}(\Omega)=-2 \pi i \left[ N_{\omega_{\lambda {\bf Q}}} \delta(\Omega-
\omega_{\lambda {\bf Q}}) \right. \nonumber \\ 
\left. + (N_{\omega_{\lambda {\bf Q}}}+1) \delta(\Omega+\omega_{\lambda {\bf Q}}) \right],
\end{eqnarray}
where $N_{\omega_{\lambda {\bf Q}}}=[e^{\omega_{\lambda {\bf Q}}/T}-1]^{-1}$ is the Planck's 
distribution function and $\omega_{\lambda {\bf Q}}$ is the phonon frequency.
To find $G^{--}$, $G^{++}$, $D^{--}$, and $D^{++}$, one uses 
\begin{eqnarray}
G^{--} = G^{-+}+G^{R},~~~G^{++} = G^{-+}-G^{A}, \nonumber \\
D^{--} = D^{-+}+D^{R},~~~D^{++} = D^{-+}-D^{A}.
\end{eqnarray}
The retarded and advanced Green's function of phonons are expressed as 
\begin{eqnarray}   
D^{R}_{\lambda {\bf Q}}(\Omega) = 
\frac{1}{\Omega-\omega_{\lambda {\bf Q}}+i0} - \frac{1}{\Omega+\omega_{\lambda {\bf Q}}+i0}, 
\end{eqnarray}
and $D^{A}_{\lambda {\bf Q}}(\Omega)=[D^{R}_{\lambda {\bf Q}}(\Omega)]^*$.
The retarded Green's function of electrons depends on the magnetic field $B$ and satisfies the equation
\begin{eqnarray}   
\left[ \varepsilon - \frac{{\bf p}^2}{2 m} + \frac{e^2 B^2}{8 m c^2} \frac{\partial^2}{\partial {\bf p}^2} 
\right] G^{R}_{\varepsilon {\bf p} t} = 1 + \frac{1}{2} \! \int \!  d \tau 
e^{i \varepsilon \tau} \! \! \int \! dt' ~~~~\\
\times \left[ \Sigma^{(im)R}_{t+\tau/2,t'} ({\bf p}) G^{R}_{t',t-\tau/2}({\bf p}) + 
G^{R}_{t+\tau/2,t'} ({\bf p}) \Sigma^{(im)R}_{t',t-\tau/2}({\bf p})  \right], \nonumber
\end{eqnarray}
where $m$ is the effective mass of electrons. The equation for the advanced Green's 
function $G^{A}$ differs from Eq. (14) only by a replacement of the superscripts $R$ 
with $A$. The integral on the right-hand side of Eq. (14) describes the influence of impurity 
scattering on the energy spectrum of 2D electrons in the magnetic field (the influence 
of other kinds of interaction on the electron spectrum is neglected here).

The electron-impurity interaction in Eqs. (3) and (7) is described by the Fourier 
transform of the random potential correlator, $w(q)$. The interaction with phonons 
[Eqs. (4) and (8)] is considered under approximations of equilibrium phonon distribution 
and bulk phonon modes. The influence of electron-phonon interaction on the phonon spectrum 
is neglected. The phonons are characterized by the mode index $\lambda$ and three-dimensional 
phonon momentum ${\bf Q}=({\bf q},q_z)$. The squared matrix element of electron-phonon 
interaction potential is represented as $M_{\lambda {\bf Q}}=C_{\lambda {\bf Q}} I(q_z)$, 
where $I(q_z)=|\left< 0|e^{iq_z z} |0 \right>|^2$ is determined by the confinement potential 
which defines the ground state of 2D electrons, $|0 \rangle$, and the function $C_{\lambda {\bf Q}}$ 
is determined by both deformation-potential and piezoelectric mechanisms of interaction:
\begin{eqnarray}
C_{\lambda {\bf Q}}=\frac{1}{2 \rho_{\scriptscriptstyle M} \omega_{\lambda {\bf Q}}} 
\biggl[ {\cal D}^2 \sum_{ij} {\rm e}_{\lambda {\bf Q} i} {\rm e}_{\lambda {\bf Q} j} Q_i Q_j  \nonumber \\
+\frac{(eh_{14})^2}{Q^4} \! \! \sum_{ijk,i'j'k'}\kappa_{ijk} \kappa_{i'j'k'} {\rm e}_{\lambda {\bf Q} k} 
{\rm e}_{\lambda {\bf Q} k'} Q_i Q_j Q_{i'} Q_{j'} \biggr].
\end{eqnarray}
Here ${\cal D}$ is the deformation potential constant, $h_{14}$ is the piezoelectric 
coupling constant, and $\rho_{\scriptscriptstyle M}$ is the material density. The sums are taken over 
Cartesian coordinate indices, ${\rm e}_{\lambda {\bf Q} i}$ are the components 
of the unit vector of the mode polarization, and the coefficient $\kappa_{ijk}$ 
is equal to unity if all the indices $i,j,k$ are different and equal to zero 
otherwise. The polarization vectors and the corresponding phonon mode frequencies 
are found from the eigenstate problem
\begin{equation}
\sum_{j} \left[K_{ij}({\bf Q})  - \delta_{ij} \rho_{\scriptscriptstyle M} \Omega^2 \right] 
{\rm e}_{\lambda {\bf Q} j}=0,
\end{equation}
where $K_{ij}({\bf Q})$ is the dynamical matrix and $\delta_{ij}$ is the Kronecker symbol. For cubic 
crystals and in the elastic approximation, the dynamical matrix is expressed through three elastic 
constants which are written in the conventional notations:
\begin{eqnarray}
K_{ij}({\bf Q}) =[(c_{11}-c_{44}) Q^2_i + c_{44} Q^2] \delta_{ij} \nonumber \\
+ (c_{12}+c_{44}) Q_i Q_j(1-\delta_{ij}). 
\end{eqnarray}
 
The 2D vector ${\bf R}_{t_1 t_2}$ standing in the exponential factors in Eqs. (3), (4),
(7), and (8) includes both the term with dc field ${\bf E}=(E_x,E_y)$ and the MW-induced 
term which depends on the frequency $\omega$ and amplitude $E_{\omega}$ of the MW field 
strength. It is convenient to use ${\bf R}^{\pm}=R_x \pm i R_y$ given by
\begin{eqnarray}   
R^{+}_{t_1 t_2}=i\frac{eE_+(t_1-t_2)}{m \omega_c} - \frac{eE_{\omega}}{\sqrt{2}m \omega} \nonumber \\
\times \left[s^*_- (e^{i \omega t_1}-e^{i \omega t_2})+s_+(e^{-i \omega t_1}-e^{-i \omega t_2}) \right],
\end{eqnarray}   
and $R^{-}_{t_1 t_2}=[R^{+}_{t_1 t_2}]^*$, where $E_{\pm}=E_x \pm i E_y$. 
The form of the complex coefficients $s_{\pm}$ depends on polarization of the radiation.
In the case of linear polarization (used for the calculations below), 
\begin{equation}
s_{\pm}=\frac{1}{\sqrt{2}} \frac{1}{\omega \pm \omega_c + i \omega_p},
\end{equation}
while in the case of circular ($-$) polarization $s_{+}=0$ and $s_{-}=1/(\omega - 
\omega_c + i \omega_p)$. Here $\omega_p$ is the radiative decay rate which determines the 
cyclotron line broadening because of electrodynamic screening effect.$^{24,25,26}$ Under 
usual experimental conditions, when the electromagnetic radiation is normally incident from 
the vacuum on the sample with dielectric permittivity $\epsilon$, and 2D electron layer 
is close to the surface of the sample, one can find $\omega_p=2 \pi e^2 n_s/ m c 
\sqrt{\epsilon^*}$, where $\sqrt{\epsilon^*}=(1+ \sqrt{\epsilon})/2$ and $n_s$ is 
the sheet electron density. It is assumed that $\omega_p$ is much larger than the 
transport scattering rate of electrons, this condition is amply satisfied
for high-mobility electrons.

In contrast to $J^{im}$ and $J^{ph}$ described above, the electron-electron collision 
integral $J^{ee}$ is not affected by the external fields. In Eqs. (5) and (9) the 
"exchange" Coulomb terms in the self-energies are neglected, since the main contribution 
to the electron-electron collision integral at low temperatures comes from small-angle 
scattering. The dynamical screening effects are also neglected. The statically-screened 
2D interaction potential is written as
\begin{equation}
U_q= \frac{2 \pi e^2}{\epsilon (q+q_0)},  
\end{equation}
where $q_0= 2 e^2 m/\epsilon$ is the inverse screening length. 

Having solved the kinetic equation (1), one can find the dissipative current density ${\bf j}$, 
averaged over the period $t_{\omega} =2 \pi/\omega$ of the microwaves:
\begin{equation}
{\bf j}= \frac{2e}{m} \int \frac{d \varepsilon}{2 \pi i} \int \frac{d {\bf p}}{(2 \pi)^2} {\bf p}~ 
\frac{1}{t_{\omega}} \int_0^{t_{\omega}} dt  G^{-+}_{\varepsilon {\bf p} t}.
\end{equation}
The current can be also expressed through the distribution function $f_{\varepsilon {\bf p} t}$ 
with the aid of Eq. (10). The factor of 2 in Eq. (21) accounts for the spin degeneracy of 
electrons (the Zeeman splitting is neglected).

\section{Calculation of resistivity}

It is assumed in the following that the temperature $T$ is small in comparison with the 
Fermi energy $\varepsilon_F$. Therefore, all the quantities whose dependence on the absolute value 
of electron momentum is slow can be taken at $p \equiv |{\bf p}|=p_F$, where $p_F$ is the 
Fermi momentum related to the sheet electron density $n_s$ as $p_F=\sqrt{2 \pi n_s}$. 
Accordingly, instead of the function $f_{\varepsilon {\bf p} t}$, one may use the distribution 
function $f_{\varepsilon \varphi t}$ which is equal to $f_{\varepsilon {\bf p} t}$ at $p=p_F$ 
and depends on the angle $\varphi$ of the quasiclassical momentum in the 2D plane.
The absolute value of the momentum ${\bf q}={\bf p}-{\bf p}'$ transferred in the elastic 
scattering by impurities is then given by $q= 2 p_F \sin(\theta/2)$, where $\theta=\varphi-\varphi'$ 
is the scattering angle. In a similar way, since the characteristic phonon energy 
$\omega_{\lambda {\bf Q}}$ is small compared to the Fermi energy, the electron-phonon 
scattering is treated in the quasielastic approximation, with $Q=\sqrt{q_z^2+4 p_F^2 
\sin^2(\theta/2)}$. The angle of the vector ${\bf q}$ is given by $\varphi_q=\pi/2+\phi$, 
where $\phi=(\varphi+\varphi')/2$.

Another approximation used below is the neglect of temporal harmonics of the distribution 
function $f_{\varepsilon \varphi t}$. Strictly speaking, excitation of such harmonics by 
microwaves can contribute to the dc current through the processes of second order in 
interaction, which leads to the so-called "photovoltaic" mechanism of MW 
photoresistance.$^{9}$ According to theoretical estimates, this contribution is not 
large, and since no experimental evidence for the importance of the photovoltaic 
mechanism has been found, its neglect is justified. In the approximations 
described above, Eq. (1) is reduced to a stationary kinetic equation for time-independent
distribution function $f_{\varepsilon \varphi}$: 
\begin{equation}
\omega_c \frac{\partial f_{\varepsilon \varphi}}{ \partial \varphi}  = J^{im}_{\varepsilon \varphi}+ 
J^{ph}_{\varepsilon \varphi} + J^{ee}_{\varepsilon \varphi}.
\end{equation}
The collision integrals standing here are calculated by using the equations of Sec. II 
(for transformation of electron-impurity collision integral, see also Refs. 9 and 27). 
This leads to the following expressions:
\begin{eqnarray}
J^{im}_{\varepsilon \varphi}=\int_0^{2 \pi} \frac{d \varphi'}{2 \pi} \nu(\theta) 
\sum_{n} [J_n (\beta)]^2 \nonumber \\ 
\times D_{\varepsilon+n \omega +\gamma} [f_{\varepsilon+n \omega +\gamma~\varphi'} - f_{\varepsilon \varphi}],
\end{eqnarray} 
\begin{eqnarray}
J^{ph}_{\varepsilon \varphi}=\int_0^{2 \pi} \frac{d \varphi'}{2 \pi} 
\sum_{\lambda,n} \int_0^{\infty} \frac{d q_z}{\pi}
m M_{\lambda {\bf Q}} [J_n (\beta)]^2 \nonumber \\
\times \left\{  
[(N_{\omega_{\lambda {\bf Q}}}+f_{\varepsilon \varphi}) 
f_{\varepsilon - \omega_{\lambda {\bf Q}} + n \omega +\gamma~\varphi'} \right. \nonumber \\
\left. - (N_{\omega_{\lambda {\bf Q}}}+1) f_{\varepsilon \varphi} ] 
D_{\varepsilon - \omega_{\lambda {\bf Q}} + n \omega +\gamma} \right. \nonumber \\
\left. + [(N_{\omega_{\lambda {\bf Q}}}+ 1 -f_{\varepsilon \varphi}) 
f_{\varepsilon + \omega_{\lambda {\bf Q}} + n \omega +\gamma~\varphi'} \right. \nonumber \\
\left. - N_{\omega_{\lambda {\bf Q}}} f_{\varepsilon \varphi} ] 
D_{\varepsilon + \omega_{\lambda {\bf Q}} + n \omega +\gamma}\right\}.
\end{eqnarray} 
The rate $\nu(\theta)$ characterizing the impurity scattering is defined as 
$\nu(\theta)=m w[2p_F\sin(\theta/2)]$, and the quantities $\omega_{\lambda {\bf Q}}$
and $M_{\lambda {\bf Q}}$ are expressed through the variables $q_z$, $\theta$, and $\phi$ 
as described above. Next, $J_n(\beta)$ is the Bessel function, and 
$D_{\varepsilon}= m^{-1} \int \frac{d {\bf p}}{(2 \pi)^2} (G^A_{\varepsilon {\bf p}}- 
G^R_{\varepsilon {\bf p}})/i$ is the density of electron states expressed in units 
$m/\pi$ (so $D_{\varepsilon}=1$ at $B=0$). In the presence of magnetic field 
($\omega_c \ll \varepsilon_F$ is assumed), the SCBA leads to the form
\begin{eqnarray}
D_{\varepsilon}= 1 + 2 \sum_{k=1}^{\infty} a_k \cos\frac{2 \pi k \varepsilon}{\omega_c}, ~~~~~~~~~ \\ 
a_k=(-1)^k k^{-1} \exp(-\pi k/\omega_c \tau_q) L_{k-1}^1 (2 \pi k/\omega_c \tau_q ), \nonumber
\end{eqnarray} 
where $\tau_q$ is the quantum lifetime of electrons and $L$ are the Laguerre 
polynomials. In the following, the case of overlapping 
Landau levels corresponding to weak magnetic fields is considered. The density 
of states is then given by the expression $D_{\varepsilon}=1-2d \cos(2 \pi \varepsilon/\omega_c)$, 
where $d = \exp(-\pi/\omega_c \tau_q)$ is the Dingle factor, and $\tau_{q}$  
is the quantum lifetime of electrons due to impurity scattering: $\tau_{q}^{-1} = 
\overline{\nu(\theta)}$ (here and below, the line over the function denotes the 
angular averaging $(2 \pi)^{-1} \int_0^{2\pi} d \theta ...$). 
The angular-dependent variables $\beta$ and $\gamma$ standing in Eqs. (23) and (24) are 
proportional to the MW and dc fields, respectively:
\begin{equation}
\beta=\frac{eE_{\omega}v_F}{\sqrt{2}\omega} 
\left|s_- (e^{i\varphi}-e^{i\varphi'}) + s_+ (e^{-i\varphi}-e^{-i\varphi'}) \right| 
\end{equation}
and
\begin{equation}
\gamma=\frac{e v_F}{2i \omega_c} [ E_{-} (e^{i\varphi}-e^{i\varphi'}) 
- E_{+} (e^{-i\varphi}-e^{-i\varphi'})],
\end{equation} 
where $v_F=p_F/m$ is the Fermi velocity.

It is convenient to expand the distribution function in the angular harmonics: 
$f_{\varepsilon \varphi}=\sum_k f_{\varepsilon k} e^{i k \varphi}$. The density 
of dissipative electric current is determined by the $k=1$ harmonic. Applying 
the usual notation $j_{\pm}=j_x \pm i j_y$, one has
\begin{equation}
j_- =\frac{e p_F}{\pi} \int d \varepsilon D_{\varepsilon} f_{\varepsilon 1}.
\end{equation}
In the regime of classically strong magnetic fields, the anisotropic ($k \neq 0$) 
part of the distribution function is expressed through the isotropic 
(angular-independent) part $f_{\varepsilon} \equiv f_{\varepsilon, k=0}$ 
directly from Eq. (22), by neglecting the angular dependence of all 
distribution functions in the collision integrals (23) and (24). The 
collision integral $J^{ee}$ does not lead to angular relaxation of the 
anisotropic distribution and, therefore, does not contribute to such a 
solution. A subsequent linearization of the collision integrals (23) 
and (24) with respect to the dc field ${\bf E}$ allows one to find a 
linear response to this field in Eq. (28). The response to $E_{-}$ defines 
the symmetric part of the diagonal conductivity, $j_- = \sigma_{d} E_-$, where
\begin{eqnarray}
\sigma_{d}=-\frac{e^2 n_s}{m \omega_c^2} 
\int_0^{2 \pi} \frac{d \theta}{2 \pi} (1-\cos \theta ) 
\int_0^{2 \pi} \frac{d \phi}{2 \pi} \sum_n [J_n (\beta)]^2 \nonumber \\
\times  \int d \varepsilon  \left[ \nu(\theta) R^{im}_{\varepsilon n} 
+ \sum_{\lambda} \int_0^{\infty} \frac{d q_z}{\pi} 
2 m M_{\lambda {\bf Q}} R^{ph}_{\varepsilon n} \right]   
\end{eqnarray} 
with 
\begin{eqnarray}
R^{im}_{\varepsilon n}= D_{\varepsilon} D_{\varepsilon+ n \omega} 
\frac{\partial f_{\varepsilon}}{\partial \varepsilon} 
+ [f_{\varepsilon}-f_{\varepsilon+ n \omega}]
\frac{\partial D_{\varepsilon}}{\partial \varepsilon} D_{\varepsilon+ n \omega} 
\end{eqnarray} 
and
\begin{eqnarray}
R^{ph}_{\varepsilon n}= D_{\varepsilon} D_{\varepsilon -\omega_{\lambda {\bf Q}} + n \omega} 
\frac{\partial f_{\varepsilon}}{\partial \varepsilon} 
(N_{\omega_{\lambda {\bf Q}}}+ 1 -f_{\varepsilon -\omega_{\lambda {\bf Q}} + n \omega}) \nonumber \\
+  [ (N_{\omega_{\lambda {\bf Q}}}+ 1) 
f_{\varepsilon} - (N_{\omega_{\lambda {\bf Q}}}+ f_{\varepsilon}) 
f_{\varepsilon -\omega_{\lambda {\bf Q}} + n \omega}] \nonumber \\
\times \frac{\partial D_{\varepsilon}}{\partial \varepsilon} D_{\varepsilon -
\omega_{\lambda {\bf Q}} + n \omega}.
\end{eqnarray} 
The resistivity $\rho$ is expressed as $\rho \simeq \sigma_d/\sigma^2_{\bot}$, where 
$\sigma_{\bot}=e^2 n_s/m \omega_c$ is the classical Hall conductivity.

In the absence of MW radiation, when $\beta =0$, one should substitute in 
Eq. (29) $[J_n (\beta)]^2=1$ for $n=0$, $[J_n (\beta)]^2=0$ for $n \neq 0$, and take 
into account that $f_{\varepsilon}$ is the equilibrium Fermi distribution function. 
Then, using $R^{im}_{\varepsilon 0}= D^2_{\varepsilon} (\partial f_{\varepsilon}/\partial 
\varepsilon)$, and $R^{ph}_{\varepsilon 0}=-(N_{\omega_{\lambda {\bf Q}}}+ 1) 
f_{\varepsilon}(1-f_{\varepsilon - \omega_{\lambda {\bf Q}}}) D_{\varepsilon} 
D_{\varepsilon - \omega_{\lambda {\bf Q}}} /T$, one obtains the expression for the 
linear resistivity of 2D systems. The phonon-assisted contribution, which is proportional 
to the factor $R^{ph}_{\varepsilon 0}$, in this case is given by Eqs. (8) and (9) of 
Ref. 18. 
 
The MW excitation leads to a significant modification of the resistivity. 
First, one should consider in Eq. (29) the terms with nonzero $n$, corresponding to 
absorption of $n$ quanta of radiation. Next, it is necessary to account for the changes 
of electron distribution function $f_{\varepsilon}$ owing to this absorption. 
This modified distribution function is found from the isotropic kinetic equation
\begin{equation} 
J^{im}_{\varepsilon}+ J^{ph}_{\varepsilon}+J^{ee}_{\varepsilon}=0.
\end{equation}
The collision integrals $J^{im}_{\varepsilon}$ and $J^{ph}_{\varepsilon}$ are given 
by Eqs. (23) and (24), where $\gamma=0$ and only the isotropic part of the distribution 
function is retained (since the anisotropic part is small, $f_{\varepsilon \varphi'} 
\simeq f_{\varepsilon \varphi} \simeq f_{\varepsilon}$). The isotropic electron-electron 
collision integral is given by$^{8,27}$
\begin{eqnarray}
J^{ee}_{\varepsilon}= \int d \varepsilon' \int d \Omega~ A_{\varepsilon \varepsilon'}(\Omega)
\left[(1-f_{\varepsilon}) (1-f_{\varepsilon'}) f_{\varepsilon+\Omega} 
f_{\varepsilon'-\Omega} \right. \nonumber \\
\left. - f_{\varepsilon} f_{\varepsilon'} (1-f_{\varepsilon+\Omega}) 
(1-f_{\varepsilon'-\Omega})  \right]D_{\varepsilon'} 
D_{\varepsilon+\Omega} D_{\varepsilon'-\Omega},~~~ 
\end{eqnarray}
and the function $A_{\varepsilon \varepsilon'}(\Omega)$ is estimated as 
$A \simeq (2 \pi \varepsilon_F)^{-1} \ln(q_0/q_1)$, where $q_1$ is a characteristic 
(small) momentum separating the ballistic and diffusive regimes of electron-electron 
scattering; see Ref. 8 for a more detailed consideration. This energy-independent 
form of $A$ is valid in the experimentally relevant region of temperatures, where 
the transferred momentum $q \sim T/v_F$ lies in the interval $q_0 \gg q \gg q_1$.

Below, only the terms linear in MW power are taken into account. Following 
the basic idea of Ref. 8, the distribution function is presented in the form
\begin{equation} 
f_{\varepsilon}=f^e_{\varepsilon} + \delta f_{\varepsilon},
\end{equation} 
where $f^e_{\varepsilon}$ slowly varies with energy on the scale of $\omega_c$, 
and $\delta f_{\varepsilon}$ oscillates with the period $\omega_c$. The function 
$f^e_{\varepsilon}$ differs from the equilibrium Fermi distribution because of 
the influence of microwaves and satisfies Eq. (32) where the oscillating 
contribution to the density of states is neglected, $D_{\varepsilon}=1$. 

One should notice that the electron-electron scattering is stronger than electron-phonon 
scattering and controls the electron distribution if the MW intensity is not high. 
Thus, the function $f^e_{\varepsilon}$ can be reasonably approximated by the heated 
Fermi distribution characterized by the electron temperature $T_e$. This temperature 
is determined from the balance equation 
\begin{equation}
W_{a}=W_{r},
\end{equation} 
where $W_{a}$ is the MW power absorbed by the electron system 
and $W_{r}$ is the power transmitted out of this system through the energy relaxation 
owing to electron-phonon scattering. The balance equation can be obtained by 
integrating the kinetic equation (32), multiplied by the energy and density of states, 
over energy $\varepsilon$. This leads to the following expressions:
\begin{equation}
W_{a}=\frac{m P_{\omega} \omega^2}{4 \pi} \left( \nu^{tr}_{im} + \nu^{a}_{ph}(\omega) \right),
\end{equation}
where $\nu^{tr}_{im} = \overline{\nu(\theta) (1-\cos\theta)}$ is the transport rate of electrons
due to electron-impurity scattering and
\begin{equation}
P_{\omega}=\left(\frac{e E_{\omega} v_F}{\hbar\omega}\right)^{2} (|s_+|^2+|s_-|^2) 
\end{equation}
is the dimensionless function proportional to MW power. The rate
\begin{eqnarray}
\nu^{a}_{ph}(\omega) = \widehat{\cal S}_1 \left\{ 
2 N_{\omega_{\lambda {\bf Q}}} - N^e_{\omega_{\lambda {\bf Q}}-\omega}
- N^e_{\omega_{\lambda {\bf Q}}+\omega} \right. \nonumber \\
+ 2 \frac{\omega_{\lambda {\bf Q}}}{\omega} 
(N^e_{\omega_{\lambda {\bf Q}}-\omega}-N^e_{\omega_{\lambda {\bf Q}}+\omega}) \nonumber \\
+ \frac{\omega_{\lambda {\bf Q}}^2}{\omega^2}(2 N^e_{\omega_{\lambda {\bf Q}}}-
\left. N^e_{\omega_{\lambda {\bf Q}}-\omega}-N^e_{\omega_{\lambda {\bf Q}}+\omega} ) \right\}
\end{eqnarray}
characterizes energy absorption due to electron-phonon scattering. The integral 
operators $\widehat{\cal S}_n$ are defined as
\begin{eqnarray}
\widehat{\cal S}_n \{ A \} \equiv \int_0^{2 \pi} \frac{d \theta}{2 \pi} 
\int_0^{2 \pi} \frac{d \phi}{2 \pi} \sum_{\lambda} \int_0^{\infty} \frac{d q_z}{\pi} \nonumber \\
\times m M_{\lambda {\bf Q}} (1-\cos \theta)^n A,
\end{eqnarray}
where $A$ is an arbitrary function which depends on the variables of integration.
The function $N^e_{\Omega}=[e^{\Omega/T_e}-1]^{-1}$ differs from the Planck's 
distribution function $N_{\Omega}$ by substitution of electron temperature $T_e$ 
in place of the lattice temperature $T$. The introduction of the operators $\widehat{\cal S}_n$ 
considerably simplifies notations in the following. The power absorbed by the lattice is 
\begin{eqnarray} 
W_{r}= \frac{m}{\pi} \widehat{\cal S}_0 \{ \omega^2_{\lambda {\bf Q}} [N^e_{\omega_{\lambda {\bf Q}}}-
N_{\omega_{\lambda {\bf Q}}}] \}.
\end{eqnarray} 
Assuming $T_e/T-1 \ll 1$, one gets $W_{r} \propto T_e-T$. According to the balance 
equation (35), this means that the heating is directly proportional to the MW 
power. Since only the contributions linear in power are considered, the relation
$T_e/T-1 \ll 1$ must be satisfied. In this approximation, the balance equation is 
rewritten as
\begin{equation}
\frac{T_e}{T}-1  = P_{\omega} \left(\frac{\omega}{2 T} \right)^2 \frac{\nu^{tr}_{im}+\nu^{a}_{ph}(\omega)}{\nu^{r}_{ph}}, 
\end{equation}
where the energy relaxation rate $\nu^{r}_{ph}$ is introduced by the expression 
\begin{equation}
\nu^{r}_{ph}=\widehat{\cal S}_0 \left\{ \frac{\omega_{\lambda {\bf Q}}}{T} 
F \left(\frac{\omega_{\lambda {\bf Q}}}{2T} \right)  \right\},
\end{equation}
and $F(x)=[x/\sinh(x)]^2$. The function $F(\omega_{\lambda {\bf Q}}/2T)$ can be also 
expressed through the Planck's distribution, $F(\omega_{\lambda {\bf Q}}/2T)=
(\omega_{\lambda {\bf Q}}/T)^2 N_{\omega_{\lambda {\bf Q}}}(N_{\omega_{\lambda {\bf Q}}}+1)$.
In the high-temperature limit, when $2 T_e \gg \omega, \omega_{\lambda {\bf Q}}$,
the dependence $W_{r} \propto T_e-T$ takes place even at high MW power. 
In this limit, $\nu^{a}_{ph}(\omega)$ becomes frequency-independent and coincides 
with the transport rate due to electron-phonon scattering. This rate is defined as 
\begin{equation}
\nu^{tr}_{ph}=\widehat{\cal S}_1 \left\{ \frac{2 T}{\omega_{\lambda {\bf Q}}} 
F \left(\frac{\omega_{\lambda {\bf Q}}}{2T} \right)  \right\}.
\end{equation}

Once the distribution function $f^e_{\varepsilon}$ is known, one can find the rapidly oscillating 
correction $\delta f_{\varepsilon}$ from the oscillating (proportional to the Dingle factors) 
part of Eq. (32). This leads to the following expression
\begin{eqnarray} 
\frac{\delta f_{\varepsilon}}{\tau_{in}} = \frac{P_{\omega}}{4} {\cal G}_{\omega} \nu^{tr}_{im} 
+ \widehat{\cal S}_0 \left\{ {\cal G}_{\omega_{\lambda {\bf Q}}} 
(N_{\omega_{\lambda {\bf Q}}}- N^e_{\omega_{\lambda {\bf Q}}}) \right\}  \nonumber \\ 
+\frac{P_{\omega}}{4} \widehat{\cal S}_1 \left\{ {\cal G}_{\omega-\omega_{\lambda {\bf Q}}} (N_{\omega_{\lambda {\bf Q}}}- N^e_{\omega_{\lambda {\bf Q}}-\omega}) +{\cal G}_{\omega+\omega_{\lambda {\bf Q}}} \right. \nonumber \\ 
\left. \times (N_{\omega_{\lambda {\bf Q}}}- N^e_{\omega_{\lambda {\bf Q}}+\omega}) 
-2 {\cal G}_{\omega_{\lambda {\bf Q}}} (N_{\omega_{\lambda {\bf Q}}}- N^e_{\omega_{\lambda {\bf Q}}}) \right\},~~
\end{eqnarray}  
where 
\begin{equation}
{\cal G}_{\omega}=\delta D_{\varepsilon-\omega}(f^e_{\varepsilon-\omega}-f^e_{\varepsilon})+
\delta D_{\varepsilon+\omega}(f^e_{\varepsilon+\omega}-f^e_{\varepsilon}),
\end{equation} 
and $\delta D_{\varepsilon}=D_{\varepsilon}-1=-2d \cos(2 \pi \varepsilon/\omega_c)$.
The inelastic scattering time $\tau_{in}$, which describes relaxation of the isotropic part 
of the distribution function due to both electron-electron and electron-phonon scattering, 
is defined as $1/\tau_{in}=1/\tau_{ee}+1/\tau_{ph}$. The electron-electron scattering 
contribution has been widely discussed in the literature$^{8,27,28}$. The 
result for the electron-phonon scattering contribution is 
\begin{eqnarray}
\frac{1}{\tau_{ph}} \simeq \widehat{\cal S}_0 \{ 2 N_{\omega_{\lambda {\bf Q}}}+1 + f^e_{\varepsilon+\omega_{\lambda {\bf Q}}} - f^e_{\varepsilon-\omega_{\lambda {\bf Q}}} \}.
\end{eqnarray} 
In derivation of this expression, the integrals $\widehat{\cal S}_0 \{ (2 N_{\omega_{\lambda {\bf Q}}}+1) 
\exp( \pm 2 \pi \omega_{\lambda {\bf Q}}/\omega_c ) \}$ containing rapidly oscillating factors are 
neglected, which allows one to consider only the outscattering contribution in the collision integral. 
At high temperatures, $2 T > \omega_{\lambda {\bf Q}}$, energy dependence of $\tau_{ph}$ becomes 
inessential. For typical parameters of GaAs quantum wells, $1/\tau_{ph}$ is small in comparison 
to $1/\tau_{ee}$, so $\tau_{in} \simeq \tau_{ee}$. 
According to Eq. (44), the function 
$\delta f_{\varepsilon}$ is composed from three terms: the first one is caused by electron-impurity 
scattering, while the two other terms are due to electron-phonon scattering. The second term does not 
oscillate with MW frequency and exists owing to electron heating. The third term is directly 
proportional to MW power and, in this sense, is analogous to the first term. Since only the effects 
linear in MW power are considered, the saturation$^{8,9}$ of $\delta f_{\varepsilon}$ 
is neglected. Strictly speaking, this approximation also requires a neglect of heating in the 
third term on the right-hand side of Eq. (44), because accounting for $T_e \neq T$ would 
produce the contributions of higher orders in MW power in this term.

Having found the distribution function, one can calculate the resistivity from Eqs. (29)-(31). 
It is written as 
\begin{equation} 
\rho=\rho_e + \delta \rho, 
\end{equation}
where $\rho_e$ and $\delta \rho$ are caused, respectively, by the components of the distribution 
function $f^e_{\varepsilon}$ and $\delta f_{\varepsilon}$. Consider first the modifications of 
resistivity associated with the function $f^e_{\varepsilon}$. The heated Fermi distribution 
$f^e_{\varepsilon}$ is substituted into Eqs. (30) and (31), and only the terms linear 
in MW power are retained ($n=0,\pm 1$). The integral over $\varepsilon$ in Eq. (29) is calculated 
under the approximation 
\begin{equation} 
2 \pi^2 T \gg \omega_c,
\end{equation} 
when the Shubnikov-de Haas oscillations are thermally averaged out. The result of this 
integration can be divided into 5 parts: 
\begin{equation}
\rho_e=\rho^{0}_{im}+ \rho^{0}_{ph} + \rho^{dis}_{im}+ \rho^{cl}_{ph}+ \rho^{dis}_{ph}. 
\end{equation}
Two of these terms come from impurity-assisted (index $im$) scattering, the first one 
is the resistivity without irradiation, and another one is a well-known$^{8,9,27,28}$ 
MW-induced correction which describes MIRO due to displacement mechanism of photoresistance:
\begin{eqnarray} 
\rho^{0}_{im}= \frac{m \nu^{tr}_{im}}{e^2 n_s } (1 + 2 d^2),~~~~~~~ \\
\rho^{dis}_{im}= -\frac{m \nu_{im}^*}{e^2 n_s} 2 d^2  P_{\omega}  \left[ \sin^2\frac{\pi \omega}{\omega_c} + 
\frac{\pi \omega}{\omega_c} \sin \frac{2\pi \omega}{\omega_c} \right],
\end{eqnarray} 
where $\nu_{im}^*=\overline{\nu(\theta)(1-\cos \theta)^2}$. This rate can be also expressed 
through the angular harmonics $\nu_k$ of the elastic scattering rate $\nu(\theta)$ as 
$\nu_{im}^*=(3 \nu_0-4 \nu_1+ \nu_2)/2$. 

The phonon-assisted (index $ph$) contribution comprises three terms. The first is described as the 
phonon-assisted resistivity modified by heating:
\begin{eqnarray} 
\rho^{0}_{ph}= \frac{2 m}{e^2 n_s} \widehat{\cal S}_1
\left\{ {\cal F}_{\lambda {\bf Q}}(0) \left(1+2d^2\cos\frac{2 \pi \omega_{\lambda {\bf Q}}}{\omega_c} \right) 
\right. \nonumber \\
\left. - 2d^2 {\cal H}_{\lambda {\bf Q}}(0) \sin \frac{2 \pi \omega_{\lambda {\bf Q}}}{\omega_c} \right \}.
\end{eqnarray} 
Here and below,
\begin{eqnarray} 
{\cal F}_{\lambda {\bf Q}}(\omega)=N_{\omega_{\lambda {\bf Q}}} - N^e_{\omega_{\lambda {\bf Q}} + \omega} 
\nonumber \\
+ \frac{\omega_{\lambda {\bf Q}} + \omega}{T_e} N^e_{\omega_{\lambda {\bf Q}} + \omega}
(N^e_{\omega_{\lambda {\bf Q}} + \omega}+1),
\end{eqnarray} 
\begin{eqnarray} 
{\cal H}_{\lambda {\bf Q}}(\omega)=(N_{\omega_{\lambda {\bf Q}}} - N^e_{\omega_{\lambda {\bf Q}}+\omega}) 
\frac{2 \pi (\omega_{\lambda {\bf Q}} + \omega)}{\omega_c}. 
\end{eqnarray} 
The term $\rho^{0}_{ph}$ does not oscillate as a function of $\omega$. In the limit $T_e=T$ it is 
reduced to the resistance $\rho_{ph}$ calculated in Ref. 18. The part of $\rho^{0}_{ph}$ proportional 
to the squared Dingle factor describes PIRO modified by heating (note that ${\cal H}_{\lambda {\bf Q}}(0)$ 
is nonzero only at $T_e \neq T$). The second phonon-assisted term 
is the MW-induced "classical" photoresistance, which is not related to quantum oscillations 
of the density of states and, therefore, does not oscillate with the magnetic field: 
\begin{eqnarray} 
\rho^{cl}_{ph}= \frac{m P_{\omega}}{2 e^2 n_s} \widehat{\cal S}_2
\left\{ {\cal F}_{\lambda {\bf Q}}(\omega) +{\cal F}_{\lambda 
{\bf Q}}(-\omega) -2{\cal F}_{\lambda {\bf Q}}(0) \right\}.
\end{eqnarray} 
This term essentially requires inelastic scattering: a formal limiting transition $\omega_{\lambda {\bf Q}} 
\rightarrow 0$ makes $\rho^{cl}_{ph}$ equal to zero. Therefore, $\rho^{cl}_{ph}$ has no analogue in the impurity-assisted tansport. The contribution (55) is strongly suppressed with decreasing frequency $\omega$ 
and increasing temperature $T$. Finally, the third phonon-assisted term can be attributed to the displacement 
mechanism of MW photoresistance: 
\begin{eqnarray} 
\rho^{dis}_{ph}= \frac{m}{e^2 n_s} d^2 P_{\omega} 
\widehat{\cal S}_2 \left\{ {\cal F}_{\lambda {\bf Q}}(-\omega)
\cos\frac{2 \pi (\omega_{\lambda {\bf Q}}-\omega)}{\omega_c} \right. \nonumber \\
+ {\cal F}_{\lambda {\bf Q}}( \omega)
\cos\frac{2 \pi (\omega_{\lambda {\bf Q}}+\omega)}{\omega_c}  
-2 {\cal F}_{\lambda {\bf Q}}(0) 
\cos\frac{2 \pi \omega_{\lambda {\bf Q}}}{\omega_c} \nonumber \\
- {\cal H}_{\lambda {\bf Q}}(- \omega) \sin \frac{2 \pi (\omega_{\lambda {\bf Q}}-\omega)}{
\omega_c} - {\cal H}_{\lambda {\bf Q}}( \omega) \nonumber \\
\left. \times \sin \frac{2 \pi (\omega_{\lambda {\bf Q}}+\omega)}{\omega_c}  
 + 2 {\cal H}_{\lambda {\bf Q}}(0)\sin \frac{2 \pi \omega_{\lambda {\bf Q}}}{\omega_c} \right\}.
\end{eqnarray} 
In contrast to the impurity-assisted contribution $\rho^{dis}_{im}$, this term contains oscillating 
functions of $(\omega_{\lambda {\bf Q}}\pm \omega)/\omega_c$ describing interference of MIRO and PIRO.

The next step is to consider the terms associated with the oscillating part of the distribution 
function, $\delta f_{\varepsilon}$, generated by microwaves. Taking into account that 
$\delta f_{\varepsilon}$ is proportional to MW power, one should retain only the 
terms with $n=0$ in Eq. (29). 
As a result,
\begin{eqnarray}
\delta \rho=-\frac{m}{e^2 n_s} \left[ \nu^{tr}_{im} \int d \varepsilon \delta R^{im}_{\varepsilon} 
+ 2\widehat{\cal S}_1 \left\{ \int d \varepsilon \delta R^{ph}_{\varepsilon} \right\} \right],   
\end{eqnarray} 
where $\delta R^{im}_{\varepsilon}$ and $\delta R^{ph}_{\varepsilon}$ are obtained by a
linearization of $R^{im}_{\varepsilon 0}$ and $R^{ph}_{\varepsilon 0}$ with respect to 
$\delta f_{\varepsilon}$:
\begin{equation}
\delta R^{im}_{\varepsilon}= D^2_{\varepsilon} \frac{\partial \delta f_{\varepsilon}}{\partial \varepsilon}, 
\end{equation} 
\begin{eqnarray}
\delta R^{ph}_{\varepsilon}= D_{\varepsilon} D_{\varepsilon -\omega_{\lambda {\bf Q}}} 
\frac{\partial \delta f_{\varepsilon}}{\partial \varepsilon} 
(N_{\omega_{\lambda {\bf Q}}}+ 1 -f^e_{\varepsilon -\omega_{\lambda {\bf Q}}}) \nonumber \\
-D_{\varepsilon} D_{\varepsilon -\omega_{\lambda {\bf Q}}} 
\frac{\partial f^e_{\varepsilon}}{\partial \varepsilon} \delta f_{\varepsilon -\omega_{\lambda {\bf Q}}}
+\frac{\partial D_{\varepsilon}}{\partial \varepsilon} D_{\varepsilon -
\omega_{\lambda {\bf Q}}} \nonumber \\
\times [ (N_{\omega_{\lambda {\bf Q}}}+ 1 - f^e_{\varepsilon -\omega_{\lambda {\bf Q}}}) 
\delta f_{\varepsilon} \nonumber \\
 - (N_{\omega_{\lambda {\bf Q}}}+ f^e_{\varepsilon}) 
\delta f_{\varepsilon -\omega_{\lambda {\bf Q}}}] .
\end{eqnarray} 
The integration over energy in Eq. (57) is done with the aid of Eq. (44) for $\delta f_{\varepsilon}$ 
and under the approximation (48). The result for $\delta \rho$ appears to be rather
complicated. It is presented below in a simplified form omitting the terms quadratic in 
electron-phonon scattering contributions: 
\begin{eqnarray} 
\delta \rho \simeq -\frac{m}{e^2 n_s} 2 d^2 \tau_{in} \nu^{tr}_{im} \left[ 
P_{\omega} \nu^{tr}_{im} \frac{2 \pi \omega}{\omega_c} \sin \frac{2 \pi \omega}{\omega_c} \right. \nonumber  \\
+ 4 \widehat{\cal S}_0 \left\{ \frac{2 \pi \omega_{\lambda {\bf Q}}}{\omega_c} \sin \frac{2 \pi \omega_{\lambda {\bf Q}}}{\omega_c} ( N_{\omega_{\lambda {\bf Q}}} -N^e_{\omega_{\lambda {\bf Q}}} ) \right\} \nonumber \\
+2 P_{\omega} \widehat{\cal S}_1 \left\{ {\cal H}_{\lambda {\bf Q}}(- \omega) \sin \frac{2 \pi (\omega_{\lambda 
{\bf Q}}-\omega)}{ \omega_c} \right. \nonumber \\
\left. + {\cal H}_{\lambda {\bf Q}}( \omega)\sin \frac{2 \pi (\omega_{\lambda {\bf Q}}+\omega)}{\omega_c}  
- 2 {\cal H}_{\lambda {\bf Q}}(0)\sin \frac{2 \pi \omega_{\lambda {\bf Q}}}{\omega_c} \right. \nonumber \\
\left. \times \cos^2\frac{\pi \omega}{\omega_c}
+ \left(2{\cal F}_{\lambda {\bf Q}}(0)-{\cal F}_{\lambda {\bf Q}}(-\omega)-{\cal F}_{\lambda 
{\bf Q}}(\omega) \right) \right. \nonumber \\
\left. \times \cos \frac{2 \pi \omega}{\omega_c} \cos^2 \frac{\pi \omega_{\lambda {\bf Q}}}{\omega_c} 
+\frac{1}{2}\left( {\cal F}_{\lambda {\bf Q}}(\omega)-{\cal F}_{\lambda {\bf Q}}(-\omega) \right)
\right. \nonumber \\
\left. \left. \times \sin \frac{2 \pi \omega}{\omega_c} \sin \frac{2\pi \omega_{\lambda {\bf Q}}}{\omega_c}  \right\} \right].
\end{eqnarray} 
This form is valid under the condition $\nu^{tr}_{im} \gg |\nu^{c1}_{ph}|$ (see the definition 
of the rate $\nu^{c1}_{ph}$ below and the corresponding discussion in the next section). The 
first term in the square brackets of Eq. (60) is the impurity-assisted scattering contribution.$^8$ 
The second one is the phonon-assisted scattering contribution which 
is caused by heating and does not oscillate with MW frequency. The third one (proportional 
to $P_{\omega}$) is the phonon-assisted scattering contribution showing interference of MIRO and PIRO. 
Since only the terms linear in MW power are considered, the functions ${\cal F}$ and 
${\cal H}$ in Eq. (60) should be taken at equilibrium, $T_e=T$. The same statement concerns Eq. (56).

The results are considerably simplified under the condition $2 T \gg \omega$. In this 
case one can write a general expression for the whole resistivity (47), which 
includes also the terms omitted previously in Eq. (60). It is convenient to 
represent the resistivity as a sum
\begin{eqnarray} 
\rho = \rho_{b} +  \rho_{os}, 
\end{eqnarray} 
where the background (non-oscillating) part is 
\begin{eqnarray} 
\rho_{b}=\frac{m}{e^2 n_s} \left[ \nu^{tr}_{im}(1+2d^2)+ \nu^{tr}_{ph} \right],
\end{eqnarray} 
and the oscillating part includes four terms:
\begin{eqnarray} 
\rho_{os}=\frac{m}{e^2 n_s} 2d^2 \biggl\{ \nu^{c1}_{ph} \nonumber \\
- P_{\omega} ( \nu^*_{im}+ \nu^{c2}_{ph}) 
\left[ \sin^2\frac{\pi \omega}{\omega_c} + 
\frac{\pi \omega}{\omega_c} \sin \frac{2\pi \omega}{\omega_c} \right]  \nonumber \\
- P_{\omega} \tau_{in} (\nu^{tr}_{im}+ \nu^{c1}_{ph})^2 \frac{2\pi \omega}{\omega_c} 
\sin\frac{2 \pi \omega}{\omega_c}  \nonumber \\
+ \left(\frac{T_e}{T}-1 \right) \left[ \nu^{s1}_{ph}+2 \nu^{s0}_{ph} \tau_{in} 
(\nu^{tr}_{im} + \nu^{c1}_{ph}) \right] \biggr\}.
\end{eqnarray} 
The phonon-assisted transport rate $\nu^{tr}_{ph}$ is given by Eq. (43), while the oscillating 
rates $\nu^{cn}_{ph}$ and $\nu^{sn}_{ph}$ are defined as
\begin{equation}
\nu^{cn}_{ph}= \widehat{\cal S}_n \left\{ \frac{2 T}{\omega_{\lambda {\bf Q}}} \cos 
\frac{2 \pi \omega_{\lambda {\bf Q}}}{\omega_c} F \left(\frac{\omega_{\lambda {\bf Q}}}{2T}\right) \right\},
\end{equation} 
\begin{equation}
\nu^{sn}_{ph}= \frac{4 \pi T}{\omega_c} \widehat{\cal S}_n \left\{ \sin 
\frac{2 \pi \omega_{\lambda {\bf Q}}}{\omega_c} F \left(\frac{\omega_{\lambda {\bf Q}}}{2T}\right) \right\}.
\end{equation} 
As seen from this definition, oscillations of $\nu^{cn}_{ph}$ and $\nu^{sn}_{ph}$ with the 
magnetic field have different phases governed by cosine (index $c$) and sine (index $s$) functions, 
respectively. The rate $\nu^{c1}_{ph}$ differs from $\nu^{tr}_{ph}$ by the presence of the oscillating 
factor $\cos (2 \pi \omega_{\lambda {\bf Q}}/\omega_c)$ under the integral operator. This rate 
describes equilibrium PIRO, the first term in Eq. (63), in the approximation of overlapping 
Landau levels ($d^2 \ll 1$) used here (notice that in Ref. 18 $\nu^{c1}_{ph}$ is denoted as 
$\nu^{(1)}_{ph}$). The direct influence of microwaves on magnetoresistance is described 
by the second and the third terms in Eq. (63), which are proportional to $P_{\omega}$. 
These terms express contributions of the displacement and inelastic mechanisms of 
photoresistance, respectively. Their frequency dependence is given by the same oscillating 
functions as in the case of impurity-assisted transport,$^{8,9,27,28}$ while the scattering 
rates are modified by electron-phonon interaction: $\nu^*_{im} \rightarrow  
\nu^*_{im}+ \nu^{c2}_{ph}$ and $\nu^{tr}_{im} \rightarrow  \nu^{tr}_{im}+ \nu^{c1}_{ph}$. 
This modification is very essential because the rates $\nu^{c2}_{ph}$ and $\nu^{c1}_{ph}$ 
oscillate with the magnetic field. The interference of phonon-induced 
magnetoresistance oscillations with MIRO is produced as a result 
of multiplication of these rates by the frequency-dependent oscillating functions.
Finally, the last term in Eq. (63) has no analogue in the theory of impurity-assisted 
magnetotransport. It is proportional to $T_e-T$ and can be described as a result 
of MW heating on phonon-assisted magnetoresistance. This term does not 
oscillate with frequency $\omega$ but it oscillates with the magnetic field, similar 
to the equilibrium magnetoresistance. The phase of these oscillations is shifted 
with respect to the phase of equilibrium PIRO. Notice that the heating factor 
$T_e/T-1$ is expressed through $P_{\omega}$ with the aid of Eq. (41), so the whole 
MW-induced correction to $\rho$ is proportional to $P_{\omega}$.

\section{Numerical results and discussion}

As follows from the above consideration, quantum oscillations of resistivity in 2D 
electron layers irradiated by microwaves demonstrate a rich and complicated behavior 
if electron-phonon interaction is taken into account. The absorption of microwaves 
in the presence of this interaction leads to the appearance of various additional oscillating 
contributions to magnetoresistance, Eq. (63). These contributions are similar 
to equilibrium PIRO in the sense that their periodicity depends on characteristic 
phonon frequencies $\omega_{\lambda {\bf Q}}$ with $Q = 2 p_F$. However, they are 
essentially non-equilibrium and some of them (those entering the last term of Eq. (63)) 
have a phase different from that of equilibrium PIRO. The non-equilibrium 
phonon-induced contributions entering the second and the third term of Eq. (63) 
show the periodicity determined by combined frequencies such as $\omega \pm \omega_{\lambda 
{\bf Q}}$ and correspond to the interference of PIRO and MIRO.

To illustrate the influence of microwaves on magnetoresistance and to compare 
contributions of different terms in Eq. (63), numerical calculations 
have been carried out. The sample parameters are taken for [001]-grown GaAs quantum 
wells of high mobility ($1.17 \times 10^{7}$ cm$^2$/V s~ at low temperature) 
experimentally investigated in Ref. 17 without MW excitation. The list of parameters 
necessary for description of electron-phonon interaction can be found in Ref. 18. 
The inelastic scattering time was estimated according to the relation$^8$ 
\begin{equation} 
\frac{1}{\tau_{in}} \simeq \lambda_{in} \frac{T^{2}}{\varepsilon_F},
\end{equation}
where $\lambda_{in}$ is a numerical constant of order unity. 
For calculations, $\lambda_{in}=1$ was chosen, which is consistent 
with experimental data.$^{29}$ The rate $\nu^*_{im}$ characterizing the displacement 
mechanism of photoresistance under impurity-assisted scattering has been calculated 
with the aid of the model of long-range random impurity potential described by the 
correlator $w(q) \propto \exp(-l_c q)$. The correlation length, $l_c \gg p_F^{-1}$, is 
determined by using the transport rate $\nu^{tr}_{im}$ found from the given mobility and  
quantum lifetime $\tau_q = 15$ ps estimated in Ref. 17.

\begin{figure}[ht]
\includegraphics[width=9.2cm]{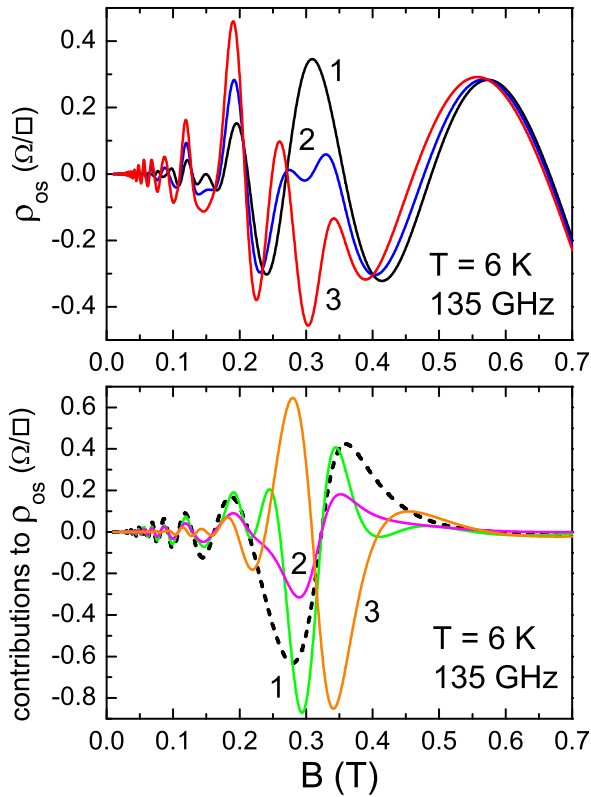}
\caption{(Color online) Upper panel. Magnetic-field dependence of the oscillating part of 
resistivity, $\rho_{os}$, calculated for the quantum well of Ref. 17 at $T=6$ K in equilibrium 
(black, curve 1) and under 135 GHz MW excitation with $E_{\omega}=2$ V/cm 
(blue, curve 2) and $E_{\omega}=3$ V/cm (red, curve 3). Lower panel. MW-induced 
contributions to $\rho_{os}$ at $E_{\omega}=3$ V/cm: displacement mechanism, second term 
in Eq. 63 (green, curve 1), inelastic mechanism, third term in Eq. 63 (magenta, curve 2), 
non-equilibrium PIRO, fourth term in Eq. 63 (orange, curve 3). The contribution due to 
elastic impurity scattering (MIRO) is shown by black dashed line.}  
\end{figure}

Figure 1 shows the oscillating part of resistivity at a lattice temperature of 6 K 
without microwaves (in equilibrium) and under 135 GHz excitation whose intensity is determined 
by the MW electric field $E_{\omega}$. For the chosen frequency, the increase in intensity 
leads to enhancement of PIRO maxima at 0.12 and 0.18 T, while the maximum at 0.3 T is 
suppressed and eventually inverted, and two additional maxima appear at its 
sides. The inversion of this PIRO peak cannot be explained as a result of simple 
superposition of phonon-induced and MW-induced oscillations and is 
attributed to interference of these kinds of oscillations, as reflected in 
Eq. (63). The lower panel of Fig. 1 shows relative contributions to the 
oscillating resistivity. Apart from the impurity-induced contribution (MIRO), this 
plot shows the total contribution of displacement mechanism, inelastic mechanism, 
and non-equilibrium PIRO. All these contributions are comparable, though the inelastic 
mechanism contribution is smaller than the others in the region around 0.3 T. 
The $B$-dependence of this contribution closely follows the MIRO plot, 
showing slight deviations caused by phonon-induced oscillations. In contrast, such 
deviations are very pronounced in the displacement mechanism contribution (see, for 
example, peak at 0.25 T and minimum at 0.4 T). The contribution of non-equilibrium 
PIRO is clearly shifted by phase with respect to the other contributions. This shift 
leads to a partial compensation of different contributions, which reduces the 
influence of microwaves on magnetoresistance.   

\begin{figure}[ht]
\includegraphics[width=9.2cm]{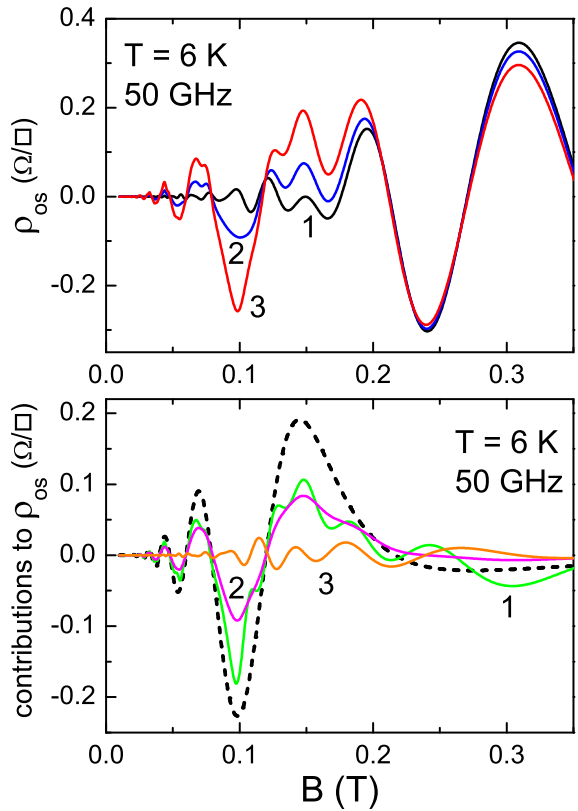}
\caption{(Color online) The same as in Fig. 1 for 50 GHz MW excitation with 
$E_{\omega}=0.5$ V/cm (blue, curve 2) and $E_{\omega}=0.8$ V/cm (red, curve 3). The MW-induced 
contributions in the lower panel are shown for $E_{\omega}=0.8$ V/cm.}  
\end{figure}

The effect of microwaves of a smaller frequency, 50 GHz, is illustrated in Fig.~2. 
The magnetoresistance is considerably changed only in the low-field region.
These changes are similar to those in Fig.~1: PIRO peaks are either enhanced or 
inverted by microwaves. Again, the inelastic mechanism contribution is weakly modified by 
electron-phonon interaction, while the displacement mechanism contribution is strongly 
modified by this interaction and shows numerous oscillations related to interference of 
PIRO with MIRO. The non-equilibrium PIRO contribution is smaller than the others. 
The relative decrease of this contribution with lowering frequency also follows 
from Eqs. (63) and (41). 

The calculations demonstrate that the displacement mechanism of MW 
photoresistance is much stronger influenced by electron-phonon interaction 
than the inelastic mechanism. This property is understood from Eq. (63) showing 
that the relative contributions caused by phonons for these two mechanisms are 
determined by the ratios $|\nu^{c2}_{ph}|/\nu^*_{im}$ and $|\nu^{c1}_{ph}|/\nu^{tr}_{im}$, 
respectively. At $T=6$ K the second ratio is small, as also seen from experimental 
data:$^{17}$ the amplitudes of equilibrium PIRO governed by the rate $\nu^{c1}_{ph}$ 
are small compared to the background resistivity. On the other hand, for long-range 
impurity potentials one has $\nu^*_{im} \ll \nu^{tr}_{im}$, while $\nu^{c2}_{ph}$ 
is approximately twice larger by amplitude than $\nu^{c1}_{ph}$, since the main 
contribution to the oscillating phonon-assisted rates $\nu^{cn}_{ph}$ comes from 
backscattering processes, $\theta \simeq \pi$. Therefore, one always has
\begin{equation} 
\frac{|\nu^{c2}_{ph}|}{\nu^*_{im}} \gg \frac{|\nu^{c1}_{ph}|}{\nu^{tr}_{im}}.
\end{equation}
With increasing temperature, when electron-phonon scattering gets stronger and 
the rates $\nu^{cn}_{ph}$ increase, the ratio $|\nu^{c2}_{ph}|/\nu^*_{im}$ becomes 
comparable to unity and the displacement mechanism contribution is considerably 
modified by this scattering. This corresponds to the temperature chosen for the 
calculations. Further increase in temperature leads to an interesting situation 
when the amplitudes of $\nu^{c2}_{ph}$ become much larger than $\nu^*_{im}$, 
so the displacement mechanism of MW photoresistance is dominated by 
electron-phonon scattering. 

It is worth noting that at low temperatures ($T \simeq 1$ K), when phonon-assisted 
contributions are yet frozen out, the inelastic mechanism is parametrically 
stronger$^{8,9}$ than the displacement one, because of large value of $\tau_{in}$. 
However, in the samples of very high mobility, like those investigated in Ref. 17, 
the contribution of the displacement mechanism experimentally proves to be important$^{12}$ 
starting from $T=2$ K. According to estimates, at this temperature the product 
$\tau_{in} \nu^{tr}_{im}$ determining the strength of inelastic mechanism is
smaller than unity. As follows from Eq. (63), the parameter $\tau_{in} \nu^{tr}_{im}$ 
also determines the relative strength of phonon-assisted contribution in the 
inelastic (third) term compared to the similar contribution in the displacement 
(second) term. Therefore, in the temperature region when phonon-induced 
oscillating contributions to resistivity become thermally activated ($T > 2$ K), 
their relative influence on the MW photoresistance via inelastic mechanism 
is already suppressed. In particular, at $T=6$ K a strong inequality, 
\begin{equation} 
\tau_{in} \nu^{tr}_{im} \ll 1,
\end{equation}
allows one to neglect the phonon-induced oscillating photoresistance due to 
inelastic mechanism. For the samples of moderate mobility ($\sim 10^6$ cm$^2$/V s), 
the applicability of Eq. (68) requires higher temperatures. Nevertheless, in the 
low-temperature region ($\tau_{in} \nu^{tr}_{im} > 1$) the ratio $|\nu^{c1}_{ph}|/\nu^{tr}_{im}$ 
remains small, and phonon-induced photoresistance due to inelastic mechanism can be 
neglected in comparison with the impurity-induced photoresistance. Notice 
also that the condition (68), being applied to non-equilibrium PIRO (the last 
term in Eq. (63)), allows one to neglect the part which comes from $\delta \rho$ and 
is proportional to $\tau_{in}$. Thus, under condition (68) one may always ignore 
the contribution to $\delta \rho$ caused by electron-phonon interaction. In contrast, 
the non-equilibrium contributions to $\rho_e$ caused by this interaction cannot be 
ignored and are important for evaluation of MW-induced magnetoresistance in 
2D systems where PIRO is observed. These contributions are described by the 
oscillating rates $\nu^{c2}_{ph}$ and $\nu^{s1}_{ph}$ in Eq. (63).

The main approximations used in the paper are discussed below in more detail.

The approximation of overlapping Landau levels corresponds to a simplified (single-mode) 
description of the oscillating density of electron states in contrast to the exact SCBA 
expression (25). Though in high-mobility samples separation of Landau levels becomes essential 
already at $B \simeq 0.1$ T, this description produces a very good agreement with the 
calculation based on Eq. (25) (see Ref. 18) for equilibrium oscillating magnetoresistance 
up to $B=0.4$ T. This means that the first PIRO harmonic dominates in the main interval 
of magnetic fields under consideration, and only the last PIRO peak at $B \simeq 0.6$ T 
appears to be slightly sensitive to the shape of the density of states. Besides, regarding 
a limited validity of the SCBA in the regime of separated Landau levels, the single-mode 
description proves to be a reasonable choice. It also provides a rigorous justification for 
introduction of inelastic scattering time necessary for analytical description of inelastic 
mechanism of photoresistance. 

The assumption of weak MW excitation means that only linear terms in the expansion of 
magnetoresistance in powers of MW intensity are taken into account. First, this 
implies a neglect of multi-photon absorption processes and formally requires that the argument 
$\beta$ of the Bessel function in Eq. (29) must be small for the scattering angles 
essential in the integration. Since the phonon-assisted scattering occurs in a broad 
range of angles, the condition $\beta \ll 1$ is equivalent to $P_{\omega} \ll 1$ and 
is satisfied by a proper choice of the MW field strength $E_{\omega}$. Next, the 
neglect of saturation effect for inelastic mechanism contribution$^{8,9}$ requires $\tau_{in} 
\nu^{tr}_{im} P_{\omega} \ll 1$. As shown above, the phonon-induced photoresistance oscillations 
in high-mobility layers are essential at $\tau_{in} \nu^{tr}_{im} < 1$, so the neglect of 
saturation is also satisfied at $P_{\omega} \ll 1$. Finally, the absence of strong heating,
$T_e/T-1 \ll 1$, is equivalent to smallness of the right-hand side of Eq. (41) and also can 
be expressed in terms of smallness of $E_{\omega}$. 

Coming back to multi-photon processes, it is worth noting that phonon-assisted multi-photon 
absorption requires much weaker MW fields than impurity-assisted multi-photon absorption. 
The reason is the long-range nature of impurity potential in high-mobility (modulation-doped) 
structures, which leads to small-angle scattering ($\theta \ll 1$) and, thereby, effectively 
smaller $\beta$ in the impurity term of Eq. (29). Since the multi-photon processes are thought 
to be responsible for the phenomenon of fractional MIRO observed at elevated MW 
intensities,$^{10,30-35}$ the above conclusion may be useful for interpretation of relevant 
experimental data. In particular, a recent experiment$^{34}$ shows several high-order fractional 
resonances in photoresistance when the MW power is still insufficient to produce these 
features as a result of multi-photon processes under electron-impurity scattering. 
A consideration of electron-phonon interaction, which enables multi-photon processes at 
weaker MW power, could possibly explain the data of Ref. 34 (indeed, the high-order 
fractional features are observed at $T = 6.5$ K, when phonon-assisted scattering may 
become important). A more detailed consideration of this problem requires a separate study 
which is beyond the scope of the present paper.


The neglect of electron-phonon and electron-electron interaction in calculation of the 
Green's function in Eq. (14) is justified at low temperatures, when the inverse quantum 
lifetime due to impurity scattering, $1/\tau_q$, is much larger than the corresponding 
inelastic scattering rates. This leads to temperature-independent density of states 
characterized by the Dingle factors used in the above calculations. However, while 
electron-phonon contribution is not essential in a wide temperature range, the contribution 
of electron-electron scattering to the Landau level broadening cannot be disregarded already 
at several Kelvin. To take this effect into account, one should replace the Dingle factors $d$ 
with temperature-dependent functions $d(T)=\exp [-\pi/\omega_c \tau_q(T)]$, where
\begin{equation} 
\frac{1}{\tau_q(T)} = \frac{1}{\tau_q} + \frac{1}{\tau_q^{ee}},~~~\frac{1}{\tau_q^{ee}}
\simeq \lambda \frac{T^{2}}{\varepsilon_F},
\end{equation}
and the numerical constant $\lambda$ (usually determined experimentally) is of the 
order of unity. The reliability of this approach is justified theoretically and proved 
in numerous works by studying temperature dependence of quantum oscillations in 
magnetoresistance.$^{12,17,28,36,37,38,39}$ The inclusion of temperature-dependent Dingle 
factors into Eq. (63) leads, in particular, to non-monotonic temperature dependence of 
PIRO amplitudes, which is also observed in experiments.$^{17,40}$

The consideration was done for the case of spatially homogeneous 2D electron system. 
The effects of a finite size of the sample, in particular, the 
influence of contacts and edges on the MW photoresistance, have been neglected. 
This includes, for instance, the neglect of confined magnetoplasmon effects in MW 
absorption.$^{41}$ The problem of magnetoresistance under inhomogeneous MW 
field distribution appears to be even more important. The observed insensitivity 
of MIRO to the sense of circular polarization,$^{42}$ the absence of MIRO in 
contactless measurenents,$^{43}$ and calculations of the field distribution in 
2D systems with metallic contacts$^{44}$ suggest possible near-contact origin of 
MIRO phenomenon in high-mobility layers. This means that, owing to large MW 
field gradients in the vicinity of contacts, there may exist an additional (caused 
by near-contact regions) oscillating contribution to resistivity$^{44}$ which exceeds 
the contribution caused by uniform MW field in the bulk of the layer.
The nature of the assumed near-contact oscillating contribution remains unclear, 
and its theoretical description is currently missing. Further investigation of this 
problem is necessary. In the meanwhile, any extension of theoretical knowledge 
about MW-induced response based on bulk properties of 2D electron gas is 
important and may help to uncover the exact nature of MIRO.  

{\it In summary}, a theoretical study of the effects of continuous MW 
irradiation on the magnetoresistance of 2D electrons interacting with impurities 
and acoustic phonons is presented. The main subject of the study was the quantum 
oscillations of resistivity caused by the scattering-assisted electron transitions 
between different Landau levels. In the absence of microwaves (in equilibrium), 
there is only one kind of such oscillations (PIRO) owing to electron-phonon scattering.
When MW radiation is applied, the transitions between Landau levels can occur 
with absorption or emission of MW quanta, either under elastic scattering by 
impurities or under inelastic scattering by acoustic phonons. The concept of microscopic 
mechanisms of MW photoresistance (displacement and inelastic mechanisms), 
previously developed for the case of electrons interacting with impurities, applies 
to both these kinds of transitions. The impurity-assisted transitions are responsible 
for the oscillations (MIRO) whose $1/B$ periodicity is determined solely by the MW 
frequency $\omega$. The phonon-assisted transitions lead to more complicated oscillations 
which are described as a result of PIRO-MIRO interference. It is demonstrated that 
such oscillations are attributed mostly to the displacement mechanism. 
Finally, heating of electron system by microwaves gives rise to an additional phonon-induced 
oscillating contribution, denoted here as non-equilibrium PIRO, which is shifted by phase 
with respect to equilibrium PIRO. This contribution is not related to specific features 
of MW excitation and appears also when electron system is heated by a dc 
field. Accounting for all the oscillating contributions leads to a peculiar 
magnetoresistance picture that should be observable in high-mobility 2D layers if the 
temperature is high enough so the electron-phonon scattering is not frozen out. Some 
relevant examples are given in Figs. 1 and 2.

One of the primary goals of this work was to emphasize the importance of electron-phonon 
interaction in description of the microwave-induced oscillating magnetoresistance. 
The author believes that the results and conclusions of this research will be useful 
for better understanding of existing experimental data and may stimulate further 
experiments and theoretical studies.\\ 

{\it Acknowledgement:} The author is grateful to I. A. Dmitriev for a helpful discussion 
of the method of moving coordinate frame.


\begin{thebibliography}{44}

\bibitem{1}
M. A. Zudov, R. R. Du, J. A. Simmons, and J. L. Reno, Phys. Rev. B
{\bf 64}, 201311(R) (2001).

\bibitem{2}
R. G. Mani, J. H. Smet, K. von Klitzing, V. Narayanamurti, W. B. Johnson,
and V. Umansky, Nature {\bf 420}, 646 (2002).

\bibitem{3}
M. A. Zudov, R. R. Du, L. N. Pfeiffer, and K. W. West, Phys. Rev. Lett. {\bf 90},
046807 (2003).

\bibitem{4}
R. L. Willett, L. N. Pfeiffer, and K. W. West, Phys. Rev. Lett. {\bf 93},
026804 (2004).

\bibitem{5}
V. I. Ryzhii, Sov. Phys. Solid State {\bf 11}, 2078 (1970); V. I. Ryzhii, R. A. Suris,
and B.S. Shchamkhalova, Sov. Phys. Semicond. {\bf 20}, 1299 (1986).

\bibitem{6}
A. C. Durst, S. Sachdev, N. Read, and S. M. Girvin, Phys. Rev. Lett {\bf 91}, 086803 (2003).

\bibitem{7}
M. G. Vavilov and I. L. Aleiner, Phys. Rev. B {\bf 69}, 035303 (2004).

\bibitem{8}
I. A. Dmitriev, M. G. Vavilov, I. L. Aleiner, A. D. Mirlin, and D.
G. Polyakov, Phys. Rev. B {\bf 71}, 115316 (2005).

\bibitem{9}
I. A. Dmitriev, A. D. Mirlin, and D. G. Polyakov, Phys. Rev. B {\bf
75}, 245320 (2007).

\bibitem{10}
S. I. Dorozhkin, J. H. Smet, K. von Klitzing, L. N. Pfeiffer, and K. W. West, 
JETP Lett. {\bf 86}, 543 (2007). 

\bibitem{11} 
S. Wiedmann, G. M. Gusev, O.E. Raichev, T. E. Lamas, A. K. Bakarov, 
and J. C. Portal, Phys. Rev. B {\bf 78}, 121301(R) (2008). 

\bibitem{12} 
A. T. Hatke, M. A. Zudov, L. N. Pfeiffer, and K. W. West, Phys. Rev. Lett. {\bf 102}, 
066804 (2009).

\bibitem{13}
M. A. Zudov, I.V. Ponomarev, A. L. Efros, R. R. Du, J. A. Simmons, and 
J. L. Reno, Phys. Rev. Lett. {\bf 86}, 3614 (2001).

\bibitem{14}
J. Zhang, S. K. Lyo, R. R. Du, J. A. Simmons, and J. L. Reno, Phys. Rev. Lett. 92, 156802 (2004).

\bibitem{15}
A. A. Bykov, A. K. Kalagin and A. K. Bakarov, JETP Lett. {\bf 81}, 523 (2005).

\bibitem{16}
W. Zhang, M. A. Zudov, L. N. Pfeiffer, and K. W. West, Phys. Rev. Lett. {\bf 100}, 036805 (2008).

\bibitem{17}
A. T. Hatke, M. A. Zudov, L. N. Pfeiffer, and K. W. West, Phys. Rev. 
Lett. {\bf 102}, 086808 (2009).

\bibitem{18}
O. E. Raichev, Phys. Rev. B, {\bf 80}, 075318 (2009).

\bibitem{19}
Y. Ma, R. Fletcher, E. Zaremba, M. D'Iorio, C. T. Foxon, and J. J.
Harris, Phys. Rev. B {\bf 43}, 9033 (1991).

\bibitem{20}
X. L. Lei and S. Y. Liu, Phys. Rev. B {\bf 72}, 075345 (2005). 

\bibitem{21}
J. Inarrea and G. Platero, Phys. Rev. B {\bf 72}, 193414 (2005). 

\bibitem{22}
X. L. Lei, J. Phys.: Condens. Matter {\bf 16}, 4045 (2004). 

\bibitem{23}
V. Ryzhii and V. Vyurkov, Phys. Rev. B {\bf 68}, 165406 (2003); V. Ryzhii, arxiv:cond-mat/0305484.

\bibitem{24}
K. W. Chiu, T. K. Lee, and J. J. Quinn, Surf. Sci. {\bf 58}, 182 (1976). 

\bibitem{25}
S. A. Mikhailov, Phys. Rev. B {\bf 70}, 165311 (2004). 

\bibitem{26}
S. A. Studenikin, M. Potemski, A. Sachrajda, M. Hilke, L. N. Pfeiffer, and K. W. West,
Phys. Rev. B {\bf 71}, 245313 (2005).

\bibitem{27}
M. Khodas and M. G. Vavilov, Phys. Rev. B {\bf 78}, 245319 (2008).

\bibitem{28}
I.A. Dmitriev, M. Khodas, A.D. Mirlin, D.G. Polyakov, M.G. Vavilov, Phys. Rev. B {\bf 80}, 
165327 (2009).

\bibitem{29}
S. Wiedmann, G. M. Gusev, O. E. Raichev, A. K. Bakarov, and J. C. Portal, 
Phys. Rev. B {\bf 81}, 085311 (2010), and references therein.

\bibitem{30}
R.G. Mani, J. H. Smet, K. von Klitzing, V. Narayanamurti, W. B. Johnson, and V. Umansky, 
Phys. Rev. Lett. {\bf 92}, 146801 (2004). 

\bibitem{31}
S. I. Dorozhkin, J. H. Smet, V. Umansky, and K. von Klitzing, Phys. Rev. B {\bf 71}, 201306(R) (2005).

\bibitem{32}
M. A. Zudov, R. R. Du, L. N. Pfeiffer, and K. W. West, Phys. Rev. B {\bf 73}, 041303(R) (2006).

\bibitem{33}
I. A. Dmitriev, A. D. Mirlin, and D. G. Polyakov, Phys. Rev. Lett. {\bf 99}, 206805 (2007). 

\bibitem{34}
S. Wiedmann, G. M. Gusev, O. E. Raichev,  A. K. Bakarov, and J. C. Portal, Phys. Rev. B, 
{\bf 80}, 035317 (2009).

\bibitem{35}
M. Khodas, H.-S. Chiang, A. T. Hatke, M. A. Zudov, M. G. Vavilov, L. N. Pfeiffer, and K. W. West, 
arxiv:cond-mat/0912.1364.

\bibitem{36}
N. C. Mamani, G. M. Gusev, T. E. Lamas, A. K. Bakarov, and O. E. Raichev,
Phys. Rev. B {\bf 77}, 205327 (2008).

\bibitem{37}
A. V. Goran, A. A. Bykov, A. I. Toropov, and S. A. Vitkalov,
Phys. Rev. B {\bf 80}, 193305 (2009).

\bibitem{38}
S. Wiedmann, N. C. Mamani, G. M. Gusev, O. E. Raichev, A. K. Bakarov, and J. C. Portal, 
Phys. Rev. B \textbf{80}, 245306 (2009).

\bibitem{39}
A. T. Hatke, M. A. Zudov, L. N. Pfeiffer, and K. W. West, Phys. Rev. B {\bf 79}, 161308(R) (2009). 

\bibitem{40}
A. A. Bykov and A. V. Goran, JETP Lett., {\bf 90}, 578 (2009).

\bibitem{41}
O. M. Fedorych, S. Moreau, M. Potemski, S. A. Studenikin, T. Saku, and Y. Hirayama, 
Intern. Journ. Mod. Phys. {\bf 23}, 2698 (2009). 
             
\bibitem{42} 
J. H. Smet, B. Gorshunov, C. Jiang, L. Pfeiffer, K. West, V. Umansky, M. Dressel, R. Meisels, 
F. Kuchar, and K. von Klitzing, Phys. Rev. Lett. {\bf 95}, 116804 (2005).
           
\bibitem{43} 
I. V. Andreev, V. M. Murav'ev, I. V. Kukushkin, J. H. Smet, K. von Klitzing, and V. Umanskii,
JETP Lett. {\bf 88}, 616 (2008).

\bibitem{44} 
S. A. Mikhailov and N. A. Savostianova, Phys. Rev. B {\bf 74}, 045325 (2006).           

 
\end{thebibliography}
\end{document}